# Results for the B-meson Decay Constant from the APE Collaboration.

C.R. Allton,[a] [*] M. Crisafulli,[a] V. Lubicz,[b] G. Martinelli,[a], F. Rapuano,[a] G. Salina,[c] and A. Vladikas[c]

[a] Dipartimento di Fisica, Università di Roma 'La Sapienza' and INFN, Sezione di Roma, Rome, Italy.
[b] Dept. of Physics, Boston Univ., Boston MA 002215, U.S.A.
[c] Dipartimento di Fisica, Università di Roma 'Tor Vergata' and INFN, Sezione di Roma II, Rome, Italy.

The decay constant for the B-meson in the static limit is calculated using the Wilson and clover actions at various lattice spacings. We show that both the contamination of our results by excited states and the effects finite lattice spacing are at most the order of the statistical uncertainties. A comparison is made of our results and those obtained in other studies. Values for $f_{B_S}^{stat}/f_B^{stat}$ and $M_{B_S} - M_B$ are also given.

## 1. Introduction

The decay constant of the $B$−meson, $f_B$, defined through the matrix element of the axial current, $A_\mu$, $< 0|A_\mu|B(p) > \equiv if_B p_\mu$ (where $B(p)$ is a $B$−meson of momentum $p$), is an essential ingredient in many calculations in the Standard Model of Particle Physics. Lattice calculations of $f_B$ in the static limit [1] can provide a phenomenologically relevant prediction so long as the lattice artefacts are kept under control. The two main sources of systematic lattice errors are those due to the effects of excited states, and those due to the finiteness of the lattice spacing, $a$. This work studies both these errors: the effects of excited states by several consistency checks; and the $O(a)$ errors by calculating $f_B^{stat}$ on lattices with various $a$ values and with both the Wilson and clover [2] action (where there are no $O(a)$ effects [3]). Both these effects are shown to be smaller than, or at most the same order as the statistical errors. Our results are in agreement with other groups which use different smearing techniques and thus have a different exposure to excited state contamination. The full results of this study will be presented elsewhere [4]

## 2. Method

$f_B^{stat}$ can be expressed as

$$f_B^{stat} = \sqrt{\frac{2}{M_B}} Z^{Ren} Z_L a^{-3/2} \qquad (1)$$

---
[*]Talk presented by C.R.Allton

where $M_B$ is the experimental value of the B-meson mass, $Z^{Ren}$ is the renormalization constant required to connect the lattice axial current in the static theory to the continuum QCD axial current, and $Z_L$ is the lattice matrix element of the static axial current between the B-meson and the vacuum. Note that in [5] the $\tilde{f}_B$ is used for $Z_L$.

Due to the well known problems of excited state contamination in local-local correlators, we are forced to use smeared interpolating operators in order to extract $Z_L$ [6]. Following [7–9] we use cubic sources for the B-meson interpolating operators, gauge fixed to the Coulomb gauge. The methods used to extract $Z_L$ are the "DD-mass", "LD-mass" and "ratio" methods detailed in [9]. Our final results are the average of the values obtained with these three methods and with the error obtained by adding the spread of the three methods in quadrature with the statistical error.

Table 1 contains the lattice parameters of our simulations. The results of the first 220 configurations for the $\beta = 6.2$ run have been published in [9]. The rest of the data presented here has not yet been published.

## 3. Checks of Method

For a given value of $a$, the main source of error in $Z_L$ is due to excited states contaminating the ground state signal [5]. In our cubic smearing method, we have two parameters to optimise in order to reduce this unwanted effect: the size of the cube, $L_S$; and the starting time of the fitting window, $t_1$. We have performed four checks to ensure that optimised values of $L_S$ and $t_1$ have been chosen, and that the excited state contributions are at most of the order of the statistical noise.

Table 1
Lattice parameters, $Z^{Ren}$ and $a$ values used together with results for $E, Z_L$ and $f_B^{stat}$. Dimensionful results are obtained from the string tension using the data in [10]. The errors are purely statistical.

| $\beta$ | Action | $N_{Configs}$ | $Z^{Ren}$ | $E(K_{crit})$ | $Z_L(K_{crit})$ | $a^{-1}$ [GeV] | $f_B^{stat}$ [MeV] | $\frac{f_{B_S}^{stat}}{f_B^{stat}}$ | $M_{B_S} - M_B$ [MeV] |
|---|---|---|---|---|---|---|---|---|---|
| 6.0 | clover | 200 | 0.81 | 0.608(8) | 0.201(7) | 1.88 | 258(9) | 1.19(3) | 79(9) |
| 6.1 | Wilson | 170 | 0.70 | 0.541(9) | 0.135(7) | 2.2 | 190(10) | 1.13(5) | 59(18) |
| 6.2 | clover | 420 | 0.81 | 0.521(9) | 0.109(6) | 2.55 | 221(12) | 1.12(3) | 58(14) |
| 6.4 | clover | 110 | 0.82 | 0.473(8) | 0.075(3) | 3.38 | 235(9) | 1.13(2) | 67(15) |
| 6.4 | Wilson | 110 | 0.72 | 0.466(9) | 0.076(5) | 3.38 | 209(14) | 1.12(2) | 56(9) |

(1) There should be plateaus in the effective masses of the local-smeared and smeared-smeared correlation functions ($C^{LD}(t)$ and $C^{DD}(t)$) for $t > t_1$. These plots have been used to determine the optimised $L_S$ and $t_1$ values for each simulation.

(2) Consider the ratio $Q(t) = R(t)/R(t+1)$, where $R(t) = C^{LD}(t)/C^{DD}(t)$. The contribution of the first excited state may be parameterised as follows: $C^{LD}(t) = Z_L Z_D e^{-Et}(1 + A_L A_D e^{-\delta t})$, and similarly for $C^{DD}(t)$ where $\delta$ is the mass gap of the excited state. For sufficiently large $t$ we have $Q(t) \approx 1 + A_D(1-e^{-\delta})(A_L - A_D)e^{-\delta t}$. Thus $Q(t) = 1$ when there is vanishing overlap between the smeared operator and the excited state (ie. when $A_D = 0$). We check that the behaviour of $Q(t)$ is consistent with unity for all our optimised values of $L_S$ and $t_1$. An example of $Q(t)$ is shown in fig.(1) for $\beta = 6.1$ with $L_S = 11$.

(3) The variation of the $Z_L$ value with $L_S$ and $t_1$ can be checked directly. For large enough $t_1$ we expect that $Z_L$ is independent of $L_S$. This check is displayed in fig.(2) where we plot the chiral value of $Z_L$ against $t_1$ for the $\beta = 6.1$ run. As $t_1$ increases, there is a clear trend for the $Z_L$ data to tend towards the same value, independent of cube size $L_S$. Similar behaviour is observed for the other runs.

(4) Once our optimal choice of $L_S$ and $t_1$ has been made and confirmed we can compare our results for $Z_L$ and the binding energy, $E$, with those from other groups working at the same pa-rameter values. This comparison is displayed in table 2 where the run at $\beta = 6.1$ with the Wilson action is compared for the three kappa values in common with the $\beta = 6.1$ simulation of [5]. The clover simulation at $\beta = 6.2$ is also compared with those of [11]. As can be seen from the table there is good agreement between our data and the data obtained with a wavefunction [5] and gauge invariant [11] smearing. However the slope of both $E$ and $Z_L$ against $\kappa$ in our data is different from [5]. To study this effect, in [4], our fitting procedure will be modified to allow the optimal values of $L_S$ and $t_1$ to depend upon $\kappa$.

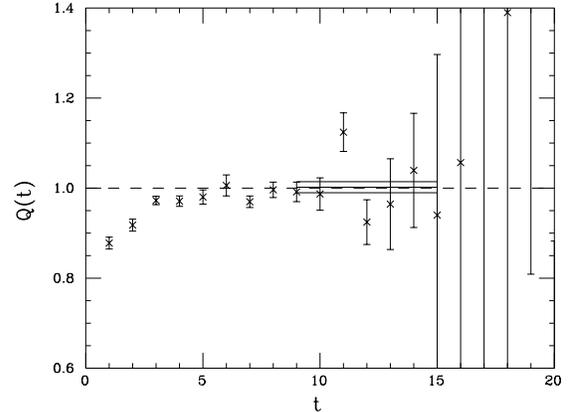

Figure 1. Ratio $Q(t)$ for the $\beta = 6.1$ run with $\kappa = 0.154$ and $L_S = 11$.

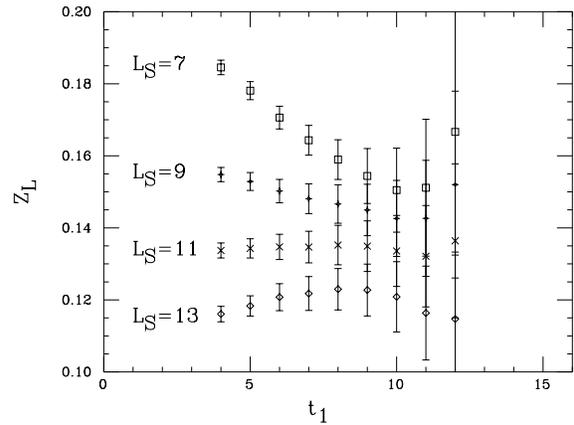

Figure 2. $Z_L$, obtained from the "LD-mass" method, against $t_1$ at $\beta = 6.1$ for various cube sizes, $L_S$, as displayed.

Table 2
Comparison with other groups.

| $\kappa$ | $E \times 10^3$ | | $Z_L \times 10^3$ | |
|---|---|---|---|---|
| | Wilson $\beta = 6.1$ | | | |
| | APE | [5] | APE | [5] |
| 0.1510 | 603(13) | 620(7) | 177(13) | 199(10) |
| 0.1540 | 552(9) | 561(11) | 142(7) | 149(12) |
| 0.1545 | 554(16) | 551(13) | 145(12) | 142(14) |
| $K_{crit}$ | 541(9) | 544(12) | 135(7) | 135(13) |
| | Clover $\beta = 6.2$ | | | |
| | APE | [11] | APE | [11] |
| 0.14144 | 561(4) | 562(3) | 134(3) | 135(4) |
| $K_{crit}$ | 521(9) | 529(5) | 109(6) | 114(4) |

## 4. Results

Once the consistency checks in section 3 have been performed and values for $Z_L$ at each $a$ has been obtained, there is left only to define $Z^{Ren}$ and $a$ in order to obtain $f_B^{stat}$ using eq.(1). The choice of $Z^{Ren}$ at finite $g^2$ [12] is beset with uncertainties, since it is known to only first order in $g^2$. There have been attempts at re-expressing this perturbative series to try and minimize unknown higher order terms by defining a new coupling $g_V^2 = g^2/u_0^4$ [13, 14]. However some ambiguity remains. It is not clear, for instance, which definition of $u_0$ to choose: eg. $1/(8K_{crit})$ or $< U_{plaq} >^{1/4}$. The difference between the two definitions can be as much as 15%. In this work we simply choose the second and await a more theoretically justified choice. The corresponding $Z^{Ren}$ values are listed in table 1.

The physical quantity used to set the scale, $a$, is again problematic. While it is certain that the string tension gives a) the most *statistically* accurate value of $a$ and b) the smallest finite lattice spacing correction (ie. it is correct to $O(a^2)$ rather than $O(a)$), its problem is that it requires model assumptions to obtain an experimental value. However, leaving aside these issues, we quote values of $f_B^{stat}$ using $a$ obtained from the string tension (using data from [10]) in table 1. From the data, within statistical errors, there is very little, if any, $a$ dependence in $f_B^{stat}$. Values for $f_{B_s}^{stat}/f_B^{stat}$ and $M_{B_s} - M_B$ are also given in table 1.

It is a pleasure to thank the APE collaboration for providing the computing resources for this study.